\documentclass[12pt,aps,prb,preprint]{revtex4}
\usepackage{latexsym}
\usepackage{amsmath, amsfonts}
\usepackage{graphicx}
\usepackage{appendix}
\usepackage{bbm}

\begin{document}

\title{Novel Shear Banding Phenomenon Probes Soap Film Friction}
\author{Ariel I. Balter}
\affiliation{Department of Physics, Indiana University}
\email{abalter@indiana.edu}
\author{Rebecca Perry}
\affiliation{Bowdoin College}
\author{James A. Glazier}
\affiliation{Department of Physics, Indiana University}
\date{\today}

\begin{abstract}
We have generated a novel form of shear banding in a 2D foam and measured the relative magnitude of drag forces on soap films at different lubrication layers.  We injected air part way along a flowing bubble field in a narrow Hele-Shaw cell. The injected air inflates bubbles as they flow by, and these bubbles form a shear band down the middle of the Hele-Shaw cell.  This channel appears to select a height that minimizes the total dissipation. Fitting data to a simple theoretical model, we show that the drag force on a soap film in contact with the wetting layer on a plate of glass is two orders of magnitude larger than the drag on a soap film in contact with another free soap film.
\end{abstract}

\maketitle

\section{Background}
Foams are studied both as unique industrial materials and also as a paradigm for complex fluids in general. In addition, many properties of foam arise from the collective effects of soap film interactions, which in turn derive from the chemistry and physics of static and flowing surfactant layers. Thus, both microscipic and macroscopic properties can be probed in foam experiments. In particular, two dimensional (2D) foams allow easy access to many properties of bubbles and foam since each bubble and each soap film interaction can be observed and analyzed \cite{Glazier1987,Debregeas2001,Kabla2003,CohenAddad2004,Cox2004,Cantat2006}. 

A series of studies by Cantat and coworkers \cite{Cantat2003,Cantat2005,Cantat2006} probes the relationship between the elastic and  rheological properties of mono- and bi-disperse foams by studying the  behavior of isolated large bubbles in a flowing monodisperse background field. Under certain conditions a large bubble will flow faster than the surrounding bubbles, pushing its way thorough the background flow. Cantat et al. explained  their observations by including the effects of both the frictional drag on individual soap films and the elastic stress in the background flow. The elastic energy ultimately drives the bubble rearrangements necessary to let the large bubble through. Debregeas et al., the first to study shear banding in foam, used Couette flow to generate shear stresses in a two dimensional foam \cite{Debregeas2001}. The shear bands in their experiments are formed and maintained by topological rearrangements at band interfaces that relax local stress. Since the shear is quantifiable, these experiments allow one to study the energy dissipated via topological rearrangements (e.g. T1 and T2) as well as their time scales.

\begin{figure}
\includegraphics[width=3.25in]{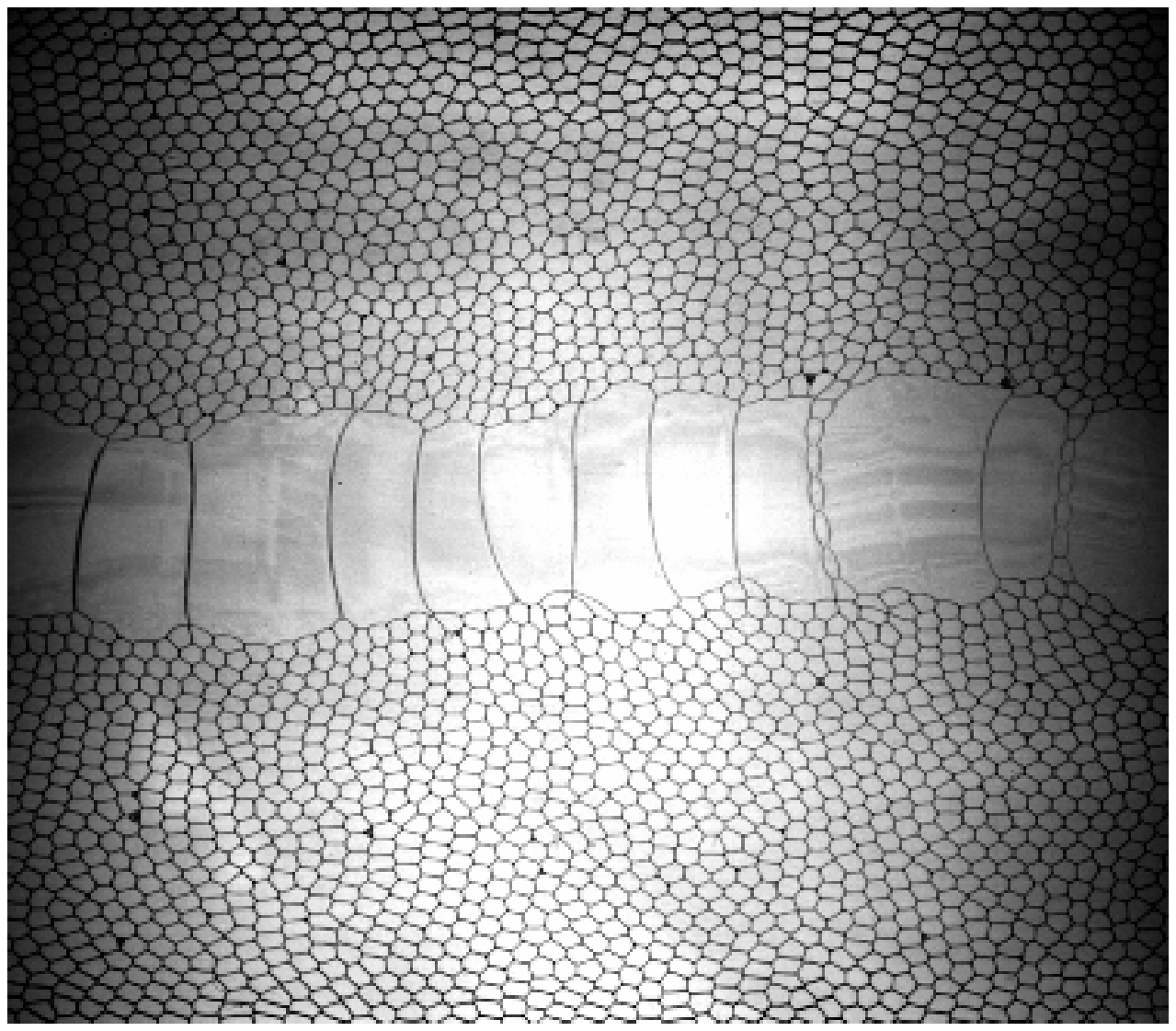}
\caption{
Typical single movie frame showing the air injection band.  Some full movies are available as additional online material.  Since the bubble generator and HSC are a closed system, all air blown into the system resides inside the bubbles.  Therefore, we can actually measure the volume of bubbles generated and the volume of injected air by measuring the flux of large and small bubbles in our movies.  We calculate the flux $\phi$, the volume of bubbles passing through a slice of the HSC per unit time, by measuring the area ($A$) and velocity ($v$) of the bubbles.  When the separation between the HSC plates is $d$, the flux is $\phi = dAv$.
}
\label{raw}
\end{figure}

\begin{figure}
\includegraphics[width=3.25in]{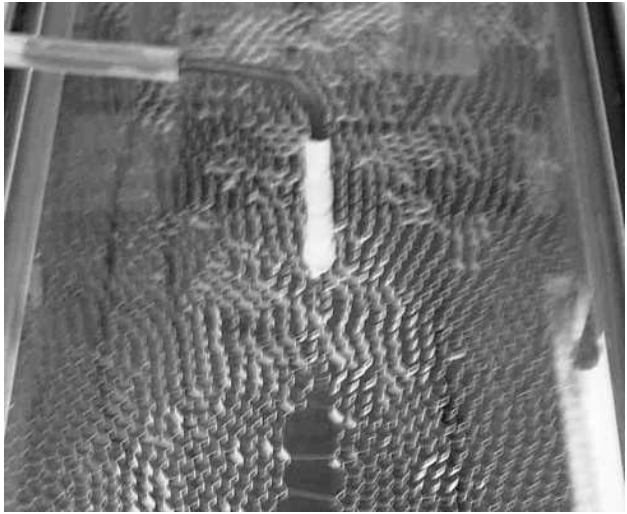}
\caption{
Air injected into the air injection port inflates which ever bubble happens to flow beneath it.  These inflated bubbles form a band of large bubbles which move in same direction as the background flow.
}
\label{injection}
\end{figure}

Unlike these previous experiments which couple frictional dissipation to topological dissipation, our configuration largely eliminates topological dissipation. We generated extended parallel shear bands by injecting air into a background flow.  The injected air parted the background flow producing three bands: two bands of background flow surrounding a central band with a different bubble structure and flow rate (figs. \ref{raw}, \ref{injection}, and supplemental material). Although severe topological rearrangements necessarily occur as the background flow is pushed aside (fig. \ref{injection}), stable coexistence of the shear bands incurs little topological cost. In a strict sense, a T1 transition occurs each time a soap film separating two large bubbles in the channel passes by a border between two small bubbles in the background flow. However, by the symmetry of the configuration, these transitions superimpose only a small periodic perturbation onto the overall dissipation.

We consistently reproduced the shear bands over a range of experimental parameters such as background flow rate, air injection rate and foam thickness. Simple geometric analysis reveals that the experimental parameters are insufficient to predict the relative size of the shear band. We hypothesize that the size emerges spontaneously to minimize drag dissipation in the system. By including both the drag between soap films and the glass plates and the drag of one soap film against another, we successfully predict the shear band height selection.  Furthermore, we obtain an estimate of the ratio of film-glass drag to film-film drag: $\sim 100$.  Although many theoretical and experimental efforts have studied the film-glass drag (see appendix A), this paper is the first, to our knowledge, to compare it to film-film drag. We hope our result leads to further experimental and theoretical study.

\section{Experimental Section}
\subsection{Method}
Our setup consists of a bubble generator and a Hele-Shaw cell (HSC). The HSC consists of two sheets of 1/2'' plate glass which we separated by one to three metal shims (0.8 mm) creating a flow cell approximately 15.5 cm wide and 71 cm long. The HSC is sealed to the bubble generator at one end and open at the other.  Approximately 30 cm from the bubble generator, the top HSC plate has a 1mm hole centered side to side. We injected air into the background flow through this port to generate the shear band. The bubble generator consists of a small tank ($\sim $1 liter) made of polycarbonate sheet. We generated bubbles by injecting air into the soap solution  (85 {\%} distilled water, 10 {\%} Joy dish soap, 5 {\%} glycerol) through a submerged manifold with four independent heads to which we can attach standard syringe needles (0.24 $mm$ i.d. to 0.58 $mm$ i.d.). The size of the bubbles each needle produces is very sensitive to the rate of air flow, so we controlled the rate of foam generation (and therefore the rate of foam flow through the cell) with the size and number of active needles rather than the rate of air. One difficulty in generating foam with multiple needles driven by the same air supply is maintaining the identical pressure drop in each needle. To handle this, we passed air from a high pressure high volume source through a fine needle valve, and then distributed this low pressure high volume air to each needle, independently, via a four-valve fish tank manifold. We also controlled the air to the air injection port with a fine needle valve.

We began each trial by adjusting needle size and flow rate to form a stable, monodisperse foam flow throughout the HSC with bubble size $\sim 3 mm$.  Then we began passing air through the air injection port. Small bubbles in the background flow inflated as they passed under the injection port, and these inflated bubbles formed a band of large bubbles flowing down the center of the HSC (figs. \ref{raw}, \ref{injection}). We took images of this channel under 45 different conditions in which we manipulated three mechanical parameters: the distance between the plates (0.8 $mm$, 1.6 $mm$, and 2.4 $mm$), the background foam flow rate (0.5 $cm^3/sec$ to 3.5 $cm^3/sec$), and the air injection rate (0 to 0.1 $cm^3/sec$).

\subsection{Analysis}
For each of the 45 configurations we recorded movies using a 1024x1044 pixel CCD camera at the fastest rate the camera would shoot ($\sim$ 30 fps) for 200 frames. We post-processed the images with Matlab scripts to track individual bubbles and calculate bubble flux. Although in the experiments we controlled the rate of bubble generation and air injection rate, we were not able to simultaneously measure these rates. However, the volume flow rate of air (both to the bubble generator and the air injection port) directly correspond to the volume flux of bubbles produced. We measured the areas ($A$) and velocities ($v$) of bubbles via image analysis, and then calculated the volume flux of bubbles $\phi = dAv$ where $d$ is the separation between the HSC plates.Because the inflated bubbles formed band of relatively stable width, one bubble wide, these "bubbles" actually took the form of  individual soap films spanning a channel between two regions of background flow. The velocity of the central band was always in the same direction as the background flow, and the speed was always equal to or greater than that of the background flow.  As in Cantat's experiments \cite{Cantat2003, Cantat2006}, below a threshold value of air injection, the central band moved in unison with the background flow (i.e. at the same speed).  Above this threshold it moved faster.  In this case, the single soap film separating large bubbles in the channel remained in contact with the soap films of the bubbles bounding the channel, though on average perpendicular to these films, as it moved along them (see fig. \ref{raw} and online movie). Intermittently, a line of bubbles perpendicular to the flow would take the place of the single soap film, however, we do not consider this as part of the phenomenon we analyzed in this study. If we injected air still faster, we passed another threshold above which the channel became unstable and fingered into the background flow.  Due to the narrow width of our channel, we have not yet investigated this fingering.

\section{Theory}
Although it would be interesting to investigate, following Cantat et al., the orgin of the threshold behavior, our chief question was how the width of the channel is determined by the parameters we experimentally controlled: channel depth (number of spacers), background foam flux (bubble generation rate) and channel bubble flux (air injection rate), size of background bubbles, and the width of the HSC. Our theoretical analysis suggested that the channel width was really only a function of the ratio of the two fluxes with a single fitting parameter, the ratio of film-film drag to film-glass drag. The control variables determine the volume flux of channel bubbles in a strictly geometric manner. But this flux can occur in a narrow, fast channel or a wide, slow one. We hypothesize  that the system selects this width in some optimal way.

We make the following simplifications: we model the drag force on an element of bubble wall as a linear response that arises independently from components of velocity proportional to and perpendicular to the element of bubble wall, and we model the bubbles in the foam as squares and rectangles with sides oriented parallel and perpendicular to the flow (fig. \ref{cartoon}). Based on these assumption, which we elaborate in appendix A, we express the power dissipated by a soap film of length $l$ parallel and perpendicular to the direction of bubble flow as $P_{\parallel} = l \eta_{\parallel}v^2$ and $P_{\bot} = l \eta_{\bot}v^2$ respectively. To calculate the power dissipated by the background flow, we consider a block of width $w = ln$ of small background flow bubbles, where $w$ is the width of the channel bubbles.  The height of the flow on either side of the channel is $(H-h)/2 = lm_S$ where $m_S$ is the number of bubbles.  Each small bubble contributes one parallel and one perpendicular film.  Considering the drag these generate, we write
\begin{equation}
\label{backgroundblock}
P=v_S^2 lmn(\eta _\parallel +\eta _\bot )
\end{equation}
for the power dissipated by this block of bubbles. The power dissipated by a large bubble is the power dissipated by the motion of a single soap film which is perpendicular the flow.  The height of the soap film is $h = lm_L$, it moves at a velocity $v_L$, and has a depth $d$, which is the distance between the plates.  The drag generated by contact with the glass plates dissipates a power
 \begin{equation}
\label{Pchannel1}
P_L =v_L^2 lm_L \eta _\bot ,
\end{equation}
while the drag generated by the relative motion of this film against the bubbles bounding the channel is
\begin{equation}
\label{Prelative}
P_{shear} =2(v_L -v_S )^2d\eta _{LS} .
\end{equation}
where $\eta_{LS}$ is the drag coefficient (per unit length) for one soap film moving across another such as at the interface where a film separating channel bubbles meets the idealized continuous film that defines the central channel. The power flux is the power dissipated per unit distance along the HSC.  We calculate this by dividing the total power by the width $w$ of our hypothetical block. Combining equations \ref{backgroundblock},\ref{Pchannel1}, \ref{Prelative} and making the associations $h=m_L l$, $H-h=2m_S l$ and $w=nl$, we express the power flux as
\begin{equation}
\label{Pflux}
P=v_S^2 \frac{H-h}{l}(\eta _\parallel +\eta _\bot )+v_L^2 \frac{h}{w}\eta
_\bot +2d\eta _{LS} \frac{1}{w}\left( {v_S -v_L } \right)^2.
\end{equation}

Since we experimentally control fluxes, not velocities, we express the power dissipation in terms of fluxes. The fluxes of small bubble flow and large bubble flow are $\phi _L =v_Lh$  and $\phi _S =v_S \left( {H-h} \right)$ respectively. The large bubble width $w$ arises from two factors: how much the area of a small bubble fills due to injection, and the selected height of the channel, $h$. Let the air be injected at a rate $\phi _A $. If the background flow of small bubbles is moving at a velocity $v_S $, then a small bubble of diameter $l$ spends about $l/v_S $ time units crossing the injection port. In doing so, it expands to an area $HW\approx l^2+\phi _A  l/v_S $. The flux of single small bubbles under the air injection point is $\phi _l =\phi _S l/H$. The total flux of channel bubbles is the sum of the flux of single small bubbles under the air injection point and the flux of injected air, i.e $\phi _L =\phi _A +\phi _l $. Thus, $hw=lH\phi _L / \phi _S$. Putting this in equation (\ref{Pflux}) yields an expression for the dissipated power flux in terms of only the experimental control parameters ($\phi _L$, $\phi _S$, $H$, and $d$), the friction, and the undetermined value of $h$:
\begin{equation}
\label{eq8}
P=\frac{\phi _S^2 }{l}\frac{\left( {\eta _\parallel +\eta _\bot }
\right)}{H-h}+\frac{\phi _L \phi _S }{lH}+\frac{2\eta _{LS} h}{lH}\frac{\phi
_S }{\phi _L }\left( {\frac{\phi _S }{H-h}-\frac{\phi _L }{h}} \right)^2
\end{equation}
We assume that the system selects the channel height $h$ in order to minimize this function. Setting $dP/dh=0$ and using the dimensionless parameters $\bar \phi  = \phi _L /\phi _S $, $\bar \eta  = (\eta _\parallel   + \eta _ \bot  )/\eta _{LS} $, $\bar h = h/H$, and $\bar d = d/H$, we obtain the solution
\begin{equation}
\label{solution}
\phi =B+\sqrt {B^2+C} ,
\end{equation}
where $B=\bar \eta {\bar d} \; {\bar h}^2/4(1-{\bar h})^2$ and $C={\bar h}^2(1+{\bar h})^2/(1-{\bar h})^2$.
We fit the experimental data to equation \ref{solution} using least squares and obtained the value $\bar \eta = 0.01$ (fig. \ref{graph}).

\section{Discussion}
We can conclude two things from the success of our model.  First, the assumption that the system optimizes the channel width to minimize the dissipation is valid.  Second, the film-film drag is much weaker than the film-glass drag.  One aspect of our analysis that lends confidence to this assertion is that our fit substantially improves when we scale the film-film drag by the length of the interface, i.e. $d$. That the film-glass drag should be larger seems reasonable since the fluid in the glass lubrication layer has a no-slip boundary condition at the glass surface, while a soap film is free to flow.  We know of no theoretical calculation or other experiment to which we can compare our observation of the film-glass drag being two orders of magnitude larger than the film-film drag. The drag on films parallel and perpendicular to the flow only enters equation \ref{solution} as the sum $\eta_{\parallel}+\eta_{\bot}$. Our success in modeling the results from this experiment suggests that future experiments could similarly measure relative drag on films parallel and perpendicular to the flow.


\section{Conclusion}
We have observed that injecting air into a steady background flow in a two dimensional foam can generate a stable shear band when the rate of injection is neither too low (in which case a band forms but moves in consort with the background foam) nor too high (in which case the shear band becomes unstable). We have observed that, for given experimental conditions, the channel chooses a unique height to minimze drag dissipation. We have formulated a simplified theory based on this hypothesis which appears to adequately explain the relationship. Our theory contains a single fitting parameter which expresses the ratio of the drag coefficient of soap film motion against the glass plates to the drag coefficient for soap film motion against another soap film: $\bar {\eta } \sim 0.01$.

\section{Acknowledgements}
Rebecca Perry participated in this project as an NSF funded REU student at Indiana University during the summer of 2006.  Ms. Perry received her B.S. from Bowdoin College in May, 2007.


\section{Appendix A}
The analysis of the flow field in a soap film which moves while in contact with the surfactant layer coating a sheet of glass is the subject of lubrication theory, and has been studied in depth \cite{Bretherton1961,Cantat2004,Denkov2006,Terriac2006}.  Well verified theory shows that the drag force should scale as the velocity to the 2/3 power.  However, to first order, reasonable agreement with experiments can often be obtained by assuming the force scales linearly with velocity \cite{Cantat2003,Cox2004}.  Evidently, for small values of the capillary number (in our case, $\mathit{C_a} \sim 10^{-2}$), the difference is not significant. Therefore, for mathematical simplicity, we use the linear relationship.

Since the drag force actually depends on fluid flow at the interface between the soap film and the lubrication layer on the glass, lubrication theory also predicts that the drag will depend on the orientation of this interface to the direction of motion.  In "viscous froth" simulations of two dimensional foam flow, Cantat et al. only applied viscous drag to the motion of soap films normal to the film \cite{Cantat2005}.  We extend this by assuming that we can decompose the drag for an interface with arbitrary orientation into components parallel and perpendicular to the soap film. Therefore, we express the force per unit length of soap film as$ \vec f_{drag} =   \left[ \eta_{\bot} {1} + (\eta_{\parallel} - \eta_{\bot}) \hat l \hat l  \right]  \vec v$. This unit length of soap film dissipates power equal to $P = \vec f_{drag} \cdot \vec v$.


\newpage

\begin{figure}
\includegraphics[width=3.25in]{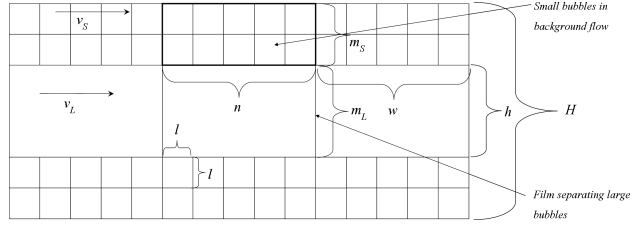}
\caption{
To estimate the power dissipated by drag, we approximate the foam as being comprised of square bubbles with sides oriented parallel and perpendicular to the direction of flow.  For simplicity, we call the bubbles in the background flow \emph{small bubbles} (S) and those in the injection band \emph{large bubbles} (L).  We begin by calculating the power dissipated by a section of flow spanning the width of a single large bubble, $w=nl$.  Then, we reduce this to a power flux (or power dissipated per unit length along the HSC) by dividing through by the width $w$.
}
\label{cartoon}
\end{figure}

\begin{figure}
\includegraphics[width=3.25in]{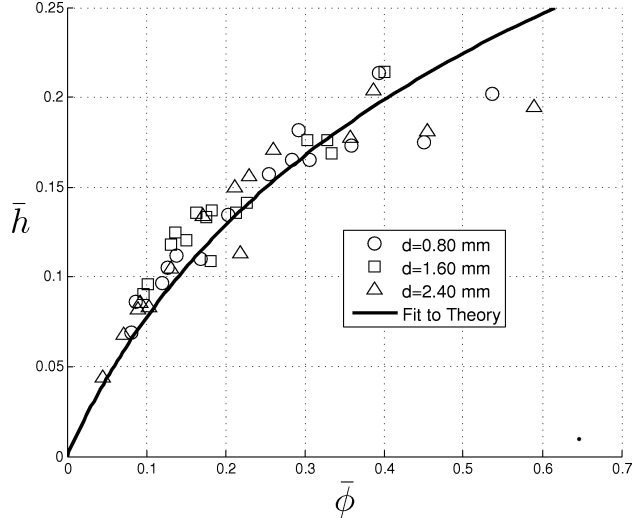}
\caption{Plot of experimentally obtained values for the normalized injection band height $\bar h=h/H$ versus the normalized bubble flux $\bar \phi = \phi_L / \phi_S$.  (See fig. \ref{cartoon} for nomenclature).  The theoretical model developed using fig. \ref{cartoon} leads to the mathematical relationship between $\bar h$ and $\bar \phi$ expressed in \ref{solution}. Adjusting the single fitting parameter $\bar \eta$ provides a good fit with $\bar \eta = 0.01$.}
\label{graph}
\end{figure}

\end{document}